\documentclass[vecphys]{svmult}

\usepackage{graphicx}        
\usepackage[bottom]{footmisc}

\begin{document}
\def\degr{\hbox{$^\circ$}}
\def\arcmin{\hbox{$^\prime$}}
\def\arcsec{\hbox{$^{\prime\prime}$}}
\def\utw{\smash{\rlap{\lower5pt\hbox{$\sim$}}}}
\def\udtw{\smash{\rlap{\lower6pt\hbox{$\approx$}}}}
\def\fd{\hbox{$.\!\!^{\rm d}$}}
\def\fh{\hbox{$.\!\!^{\rm h}$}}
\def\fm{\hbox{$.\!\!^{\rm m}$}}
\def\fs{\hbox{$.\!\!^{\rm s}$}}
\def\fdg{\hbox{$.\!\!^\circ$}}
\def\farcm{\hbox{$.\mkern-4mu^\prime$}}
\def\farcs{\hbox{$.\!\!^{\prime\prime}$}}

\title*{Symmetry and Asymmetry in "born again" Planetary Nebulae}
\author{S. Kimeswenger\inst{1}\and
A.A. Zijlstra\inst{2}\and P.A.M. van Hoof\inst{3}\and M.
Hajduk\inst{4}\and M.F.M. Lechner\inst{1}\and G.C. Van de
Steene\inst{3}\and K. Gesicki\inst{4} }
\institute{Institute of Astro- \& Particle Physics, University
Innsbruck, Technikerstr. 25, 6020 Innsbruck, Austria
\protect\newline\texttt{Stefan.Kimeswenger@uibk.ac.at} \and
University of Manchester, School of Physics \& Astronomy, PO Box 88,
Manchester M60 1QD, GB\and Royal Observatory of Belgium, Ringlaan 3,
Brussels, Belgium \and Centrum Astronomii UMK, ul. Gagarina 11,
PL-87-100 Torun, Poland}
%
%
\authorrunning{S. Kimeswenger, et al.}
\titlerunning{Symmetry and Asymmetry in "born again" PNe}
\maketitle

\begin{abstract}
While in the past spheroidicity was assumed, and still is used in
modeling of most nebulae, we know now that only a small number of
planetary nebulae (PNe) are really spherical or at least nearly
round. Round planetary nebulae are the minority of objects. In case
of those objects that underwent a very late helium flash (called
VLTP or "born-again" PNe) it seems to be different. The first,
hydrogen rich PN, is more or less round. The ejecta from the VLTP
event is extremely asymmetrically. Angular momentum is mostly
assumed to be the main reason for the asymmetry in PNe. Thus we have
to find processes either changing their behavior within a few
hundred to a few thousands of years or change their properties
dramatically due to the variation of the abundance. They most likely
have a strong link or dependency with the abundance of the ejecta.

\end{abstract}

\section{The "Family"}
\label{kimeswenger:sec:1}

Up to now it is under discussion, which objects are member of the
"family". Hydrogen poor ejecta and knots are known in several PNe.
Only a few of those have a clear signature of a VLTP event like it
is described for the first time in \cite{vltp83}. Hot PG\ 1159 white
dwarfs (so called after the prototype PG\ 1159--035 who has no
surrounding PN) are also thought to be "remnants" of this kind of
event. But less than half of them has a PN - although most of them
are hot an luminous enough to be still in the evolutionary stage
predicting one. Most of those having a PN (\cite{GoTy00} list 65
hydrogen poor central stars of PNe) do not show (prominent) hydrogen
poor ejecta (e.g. the hottest known star of that class -
RXJ2117.1+3412 \cite{werner07}). Thus, selecting only those having a
hydrogen rich "normal" PN and prominent hydrogen poor ejecta in the
core - and if it is observable - a hydrogen poor central star
(CSPN), the family contains the following objects (galactic
positions according to \cite{ks2001}) . They are sorted by the age
of the VLTP event.

\subsection{A30 (GPN G208.55+33.28)}
\label{kimeswenger:sec:1.1}

This object is most likely the oldest and best studied member of the
family. After \cite{Jacoby_Ford81} and \cite{Jacoby_Ford83}
discovered the unusual abundance of the central knots, a wide
variety of wavelengths were used to derive physical properties. It
shows all features of a "born-again" PNe
\begin{itemize}
\item A hydrogen poor central star with prominent C and O wind emission
lines and unusual wind properties (\cite{kaler84}, \cite{wind1}).
\item Fast moving hydrogen poor knots in a normal old PN
(\cite{Jacoby_Ford83}, \cite{reay83}).
\item Unusual hot small carbonous dust grains near the core in a
ring/belt like structure (\cite{dust1}, \cite{dust2})
\end{itemize}

\vspace{-3mm}

\begin{figure}
\centering
\includegraphics[width=9.5cm]{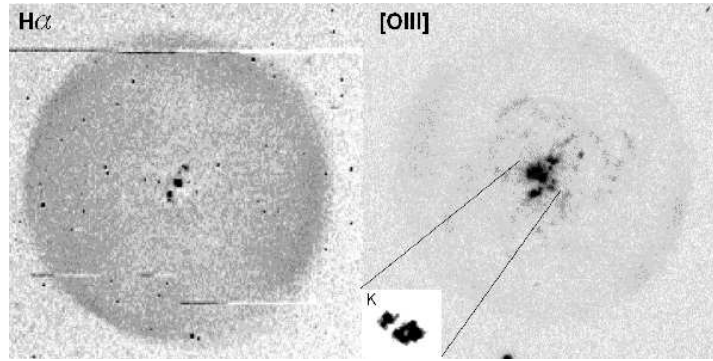}
\caption{Abell 30: The H$\alpha$ image from \cite{balick} and the
[OIII] image from the ING archive show the perfect shape of the
hydrogen rich old PN and the clumpy ejecta. The insert shows the K
band image from \cite{ks_97_a30}. It pronounces a "belt" of hot
dusty material perpendicular to the main axis defined by the VLTP
ejecta.} \label{kimeswenger:fig:a30}
\end{figure}


It is more or less the "prototype" for the class. The old nebula is
the "perfect" example of a round PN. As A30 is high above the
galactic plane, shaping by pressure equilibrium with the ISM can't
be the prominent mechanism for the perfect geometry here.

\subsection{A78 (GPN G081.29-14.91)}
\label{kimeswenger:sec:1.2}

\begin{figure}
\centering
\includegraphics[width=9.5cm]{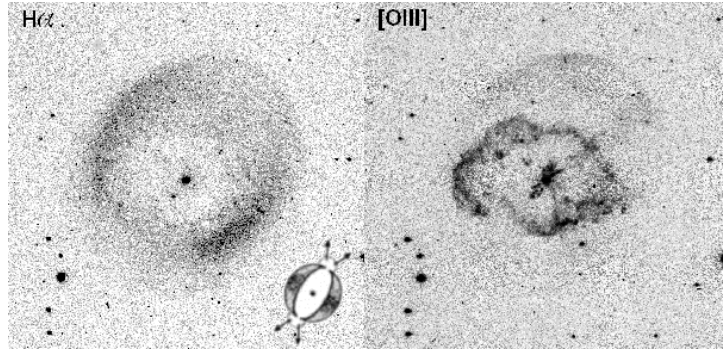}
\caption{Abell 78: The H$\alpha$ images (Calar Alto) show a barrel
type structure (like that shown in the sketch by \cite{balick}). The
[OIII] image shows the clumpy ejecta. } \label{kimeswenger:fig:a78}
\end{figure}

\vspace{-2mm}

This old nebula seems to be shaped like a barrel, inclined to the
line of sight. But this view is enhanced by an excitation effect.
[OIII] is more prominent at the poles - there more UV radiation
leaks through the ringlike structure of the inner VLTP material. As
\cite{ks_98} show in their ISO study it has, similar to A30, a
"belt" of dusty material perpendicular to the main direction of the
fast ejecta. Outside an inner knotty bipolar ejecta a wide clumpy
ring nearly leaving the area of the old PN is obvious. Thus also
here the old nebula is by far more spherical than the newly formed
VLTP material.

\subsection{IRAS 15154-5258 (GPN G324.08+03.53)}
\label{kimeswenger:sec:1.3}

Both the old nebula (diameter 32") as well as the newly formed
ejecta (7") are perfectly circular (using only the outer boundary
for classification).
\begin{figure}[ht]
\centering
\includegraphics[width=7.0cm]{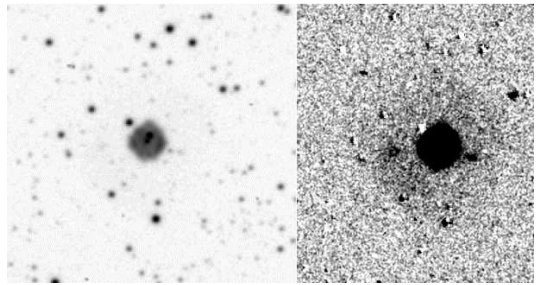}
\caption{IRAS 15154-5258: The 1991 [OIII] image from the ESO archive
(upper) show the hydrogen poor nebula (7") and the very faint older
smooth round nebula (32"). } \label{kimeswenger:fig:iras}
\end{figure}
But the inner nebula shows many radial structures. Additionally in
the HST image taken with filter F656N it has a well pronounced
bipolarity with two, possibly jet like, extensions along the axis.
Although this filter is centered at H$\alpha$ the spectrum by
\cite{manchado89} (it was integrated over 27" thus both nebulae and
still [NII] is more prominent than H$\alpha$) shows that a major
fraction of the radiation most likely originates from the [NII]
lines (still waiting for verification). An expansion velocity s not
know for this object. Nearly perpendicular to the optical "symmetry"
axis the MSX C, D \& E Band sources are extended and thus mark a
possible dust belt. But the detections are near the limit.

\begin{figure}[ht]
\centering
\includegraphics[width=7.0cm]{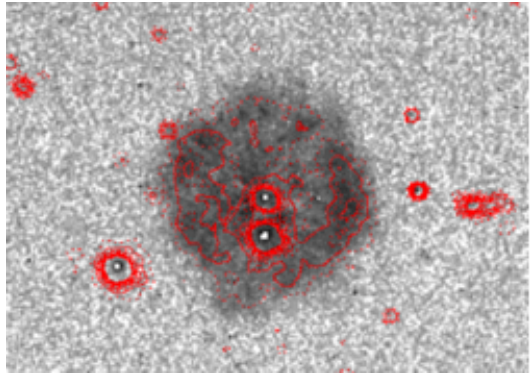}
\caption{IRAS 15154-5258: The HST images (lower: grey scale =
[OIII], contours = [NII]+H$\alpha$) show clumpy ejecta and even a
jet like structure in the same direction like the main "symmetry" of
the inner contours. } \label{kimeswenger:fig:irasB}
\end{figure}

\vspace{-2mm}

\subsection{CK Vul (NOVA Vul 1670)}
\label{kimeswenger:sec:1.4}

CK Vul is discussed controversially. The images by \cite{naylor} and
\cite{Hajduk2007} show the unusual structure of this object. While
the light curve of the 1670 eruption and the recent studies of the
knots 4 and 5 in the ejecta by \cite{Hajduk2007} fit well to a VLTP
scenario, the low luminosity of the re-ionized core found in their
radio observations does not match at all. The object consists of a
large nebula shaped like an eight and fast moving hydrogen poor
knots. The proper motion (derived from narrow band images taken 1991
and 2004) show no change at the main structure (slow moving ?) while
the knots clearly originate from the 1670 event and point exactly
towards the newly found obscured radio source. If it belongs to the
"family" of VLTP objects it is clearly that one nearest to the
galactic plane and with a highly asymmetric first PN. It is also
outstanding with respect to the fact, that the hydrogen poor knots
already are at the boundary of the other nebulosity. This would
imply a very small, and thus very young, first PN - but then a VLTP
should not occur (\cite{herwig}).


\subsection{A58 (GPN G037.60-05.16) \& its central star V605 Aql}
\label{kimeswenger:sec:1.5}

\begin{figure}[ht]
\centering
\includegraphics[width=8.2cm]{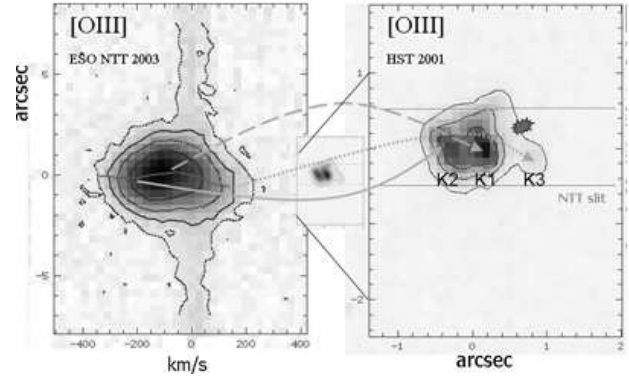}
\caption{V605 Aql: The NTT spectrum (left) and the HST image right.
The small insert gives the HST image at the same spatial scale like
the spectrum for comparison. The suspected center lies just above
K3.
 } \label{kimeswenger:fig:v605}
\end{figure}

A\ 58 was discovered, after it was related to a thermal pulse
scenario already by \cite{WoFa73}, to have a hydrogen poor central
knot by \cite{Seitter}. \cite{Pollacco1992} obtained high resolution
spectra of the central region. As we know now from HST imaging the
"standard" slit position E-W wasn't perfect to dissemble the
asymmetry. On the other hand they had a very wide slit (2"). They
derived, without knowing the orientation of the central knots,
spectra from K1 (see figure \ref{kimeswenger:fig:v605}).

\begin{figure}
\centering
\includegraphics[width=9cm]{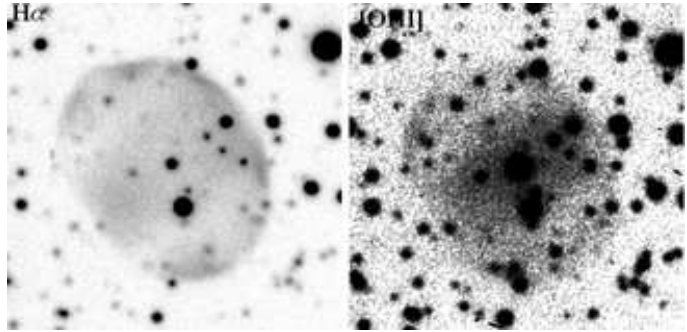}
\caption{A58: The old nebula again is very roundish with some weak
ISM interaction at one side The slightly enhanced elongated inner
[OIII] part is orientated along the axis trough K1, K2 and the
suspected center (see figure below).
 } \label{kimeswenger:fig:a58}
\end{figure}

During an 2003 ESO NTT run KS obtained medium resolution spectra
positioned along and perpendicular to the directions known now. This
shows that the knots K1 and K2 are most likely not symmetrically
around the central source. As both are blueshifted they are coming
towards us and only K3 is a wing of straylight material coming from
the redshifted side around a highly obscured dusty central region.
Assuming such a geometry, a suspected center is 0\farcs8
($\Delta\alpha = - 0\farcs72; \Delta\delta = - 0\farcs34$) from the
gaussian center defined by the emission of the (bright) knots.

\newpage
The coordinates of the Gaussian (dominated by the flux of K1 \& K2)
centered on the knots on the HST images, re-calibrated with UCAC2
sources, and those of the expected real center are:
\begin{center}
\begin{tabular}{rcrcrcr}
\multicolumn{3}{c}{[NII] \& [OIII] knots} &$\qquad\qquad$& \multicolumn{3}{c}{suspected center} \\

$\alpha_{\rm ICRS2000}$ &=& 19$^{\rm h}$18$^{\rm h}$20\fs55 & &
$\alpha_{\rm ICRS2000}$ &=& 19$^{\rm h}$18$^{\rm h}$20\fs50 \\
$\delta_{\rm ICRS2000}$ &=& +1\degr46\arcmin59\farcs25 & &
$\delta_{\rm ICRS2000}$ &=& +1\degr46\arcmin58\farcs92 \\
\end{tabular}
\end{center}

\subsection{GPN G010.47+04.41 \& its central star V4334 Sgr ($\equiv$ Sakurai)}

The old PN around V4334 Sgr is clearly a round PN with a halo. Its
morphology shows striking similarities to classical halo PNe like
NGC 2438. Such PNe are typically near the turnaround of the
evolution in the T vs. L diagram at maximum temperature and near
maximum luminosity. This is a little bit too early in the
evolutionary tracks for VLTP events.

\begin{figure}[ht]
\centering
\includegraphics[width=9cm]{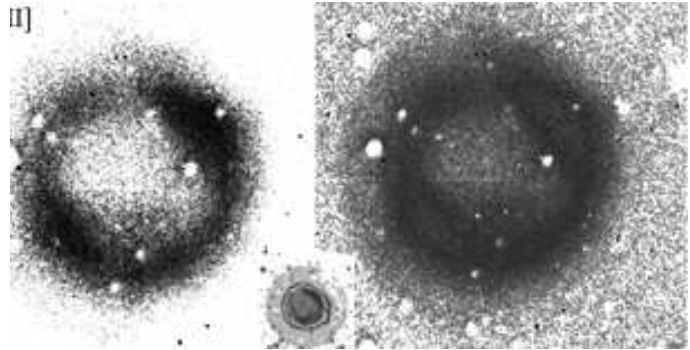}
\caption{Sakurai: The old round nebula is a halo PN like NGC 2438
(small insert).
 } \label{kimeswenger:fig:sakurai}
\end{figure}

\begin{figure}[ht]
{\includegraphics[width=3.9cm]{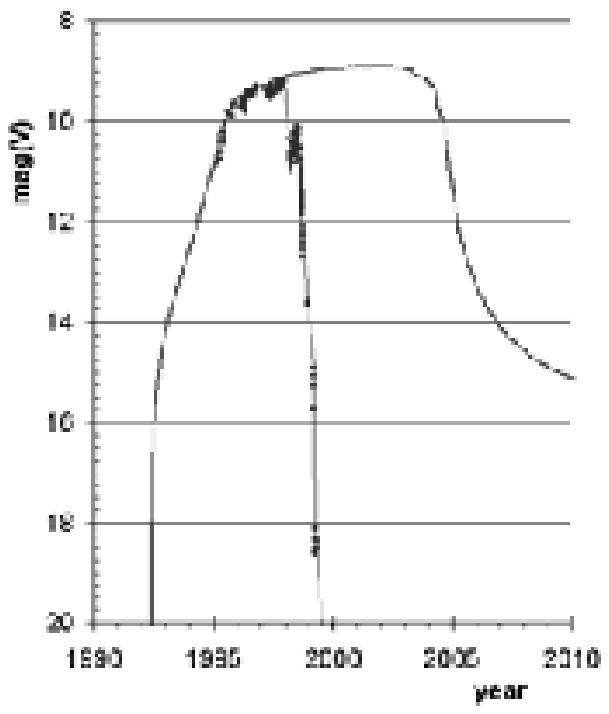}
\includegraphics[width=7.7cm]{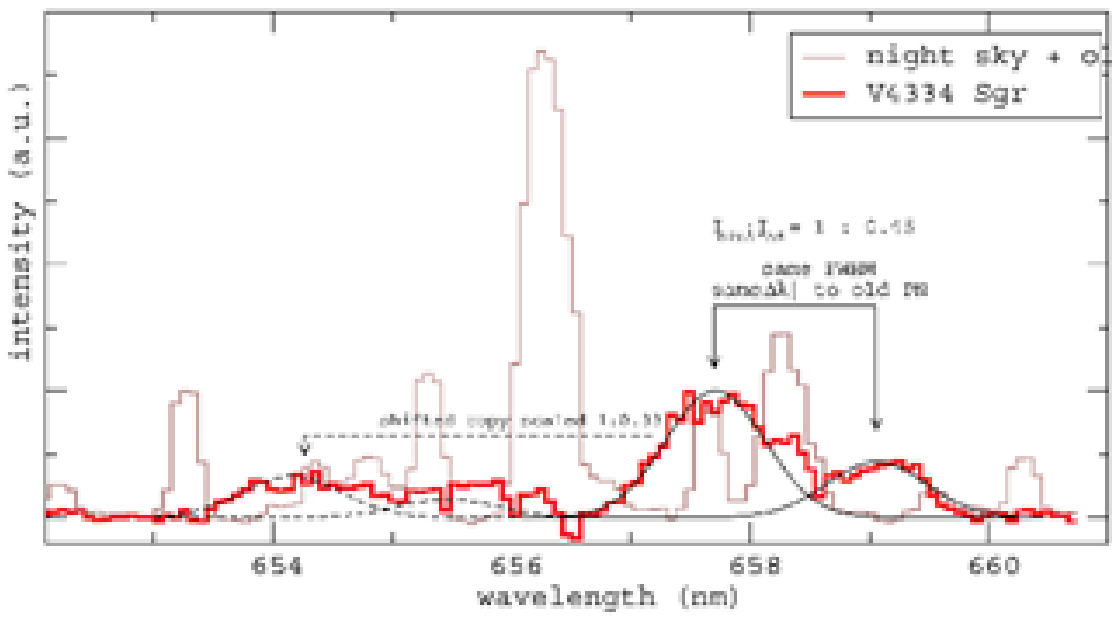}}
\caption{V4334 Sgr: The photometry (left) of the born-again core.
The data points superimposed to the prediction by the evolutionary
track my \cite{herwig} shows an obscuration by a factor of $\gg
10^4$ when the re-ionization started. The spectrum of the core (ESO
VLT FORS) shows an obscuration of the knots on the far (red) side by
only a factor of 2. The subtracted sky+old PN is indicated to show
the problem of identification of a (possible) Hydrogen or Helium
component in these spectra.
 } \label{kimeswenger:fig:V4334}
\end{figure}

The recent observations of onset of photo-ionization is presented in
detail in the contribution by PvH in this issue. The newly formed
core is too young to be resolved yet. The asymmetry of the radio
core stated by \cite{Science} was not confirmed by recent VLTI
observations (\cite{SakLetter}). We know from optical observations
that the central star is highly obscured by dust (see Fig.
\ref{kimeswenger:fig:V4334}). But ratio of the forward and backward
clumps in the [NII] lines of the newly formed core do not show such
an extreme ratio. This can be caused by clumps along an inclined
axis only - similar to the situation in V605 Aql.

\newpage
\relax
\section{The Imagination}

Commonly it is believed that asymmetry has its origin in transfer of
angular momentum. The two events - the first regular PN and the VLTP
ejecta are following each other in a few thousand to ten thousand
years only.
\begin{figure}[ht] \centering
\includegraphics[height=5cm]{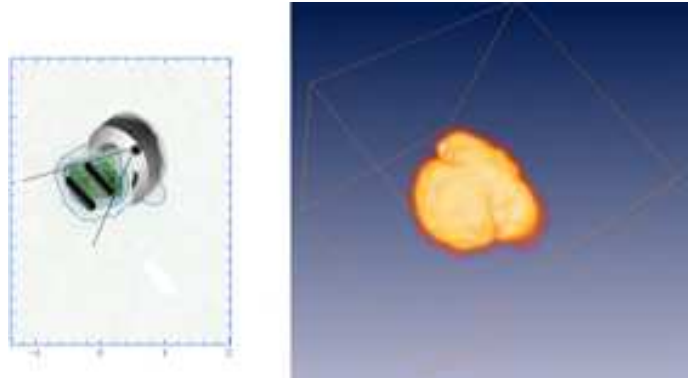}
\caption{A sketch of the born again core superimposed on the
A58/V605 Aql image (left) and the result of a hydrodynamic
simulation with a massive thick wind from a slowly rotating star. }
\label{kimeswenger:fig:model}
\end{figure}
A binary system or even  orbits of massive planets do not change
significantly during these phases. We are searching for a mechanism
that {\bf increases} its efficiency significantly from the one to
the other event. As there is no obvious mechanism changing orbits,
we have to think about mechanisms related to abundances. The
efficiency forming dust and thus a dusty torus increases
dramatically due to the carbon rich nature after the VLTP. We
believe, that there lies the key difference. It has to happen near
or at the surface of the central star. As shown by \cite{speck}, the
change of effective gravity as given by
$$ \alpha = 1 - {\omega^2 R_*^3 \over G M_*}$$
changes the mass loss around the equator. In case of a carbonous
chemistry the dust formation grows extraordinarily and builds a
waist belt - especially during the AGB like phase just after the
VLTP event. As shown by Icke (this volume) such a dusty belt leads
to colliding counter-shocks of the later incipient thin hot wind
forming a jet-like structure. The first model calculations look
promising - a slowly rotating star builds such a structure in case
of this special kind of wind.\\On the other hand we do not see an
obvious reason, why the number of round PNe (for the original first
mass loss) dominates in case of the VLTP objects. One only can
speculate; VLTP events, like those observed here, occur only in slow
or non rotating stars ?

%
%
%

\end{document}